\documentclass[submission,PhysLectNotes]{SciPost}
\usepackage{amsfonts,amssymb,mathtools}
\usepackage{bm}
\usepackage{graphicx,color,soul,caption,subcaption}
\usepackage[normalem]{ulem}

\hypersetup{
    colorlinks,
    linkcolor={red!50!black},
    citecolor={blue!50!black},
    urlcolor={blue!80!black}
}

\usepackage[bitstream-charter]{mathdesign}
\DeclareSymbolFont{usualmathcal}{OMS}{cmsy}{m}{n}
\DeclareSymbolFontAlphabet{\mathcal}{usualmathcal}

\newcommand{\dd}{\mathrm{d}}

\newcommand{\mean}[1]{\langle #1 \rangle}
\newcommand{\Int}[1]{\int\dd #1\;}
\newcommand{\IInt}[3]{\int_{#2}^{#3}\dd #1\;}

\renewcommand{\vec}[1]{\mathbf #1}

\newcommand{\al}{\alpha}

\newcommand{\Gam}{\Gamma}
\newcommand{\eps}{\varepsilon}
\newcommand{\kap}{\kappa}
\newcommand{\lam}{\lambda}
\newcommand{\Lam}{\Lambda}

\newcommand{\om}{\omega}
\newcommand{\Om}{\Omega}

\newcommand{\x}{\vec r}

\newcommand{\im}{\text{i}}

\begin{document}
\sloppy
\begin{center}{\Large \textbf{
  Perturbative dynamic renormalization of scalar field theories in statistical physics \\
}}\end{center}

\begin{center}
  Nikos Papanikolaou and Thomas Speck
\end{center}

\begin{center}
  Institute for Theoretical Physics IV, University of Stuttgart, 70569 Stuttgart, Germany
  {\small \sf thomas.speck@itp4.uni-stuttgart.de}
\end{center}

\section*{Abstract}
{\bf Renormalization is a powerful technique in statistical physics to extract the large-scale behavior of interacting many-body models. These notes aim to give an introduction to perturbative methods that operate on the level of the stochastic evolution equation for a scalar field (e.g., density), including systems that are driven away from equilibrium and thus lack a free energy. While there is a large number of reviews and lecture notes, many are somewhat scarce on technical details and written in the language of quantum field theory, which can be more confusing than helpful. Here we attempt a minimal and concise yet pedagogical introduction to dynamic renormalization in the language of statistical physics with a strong focus on how to actually perform calculations. We provide a symbolic algebra implementation of the discussed techniques including Jupyter notebooks of two illustrations: the KPZ equation and a neural network model.}

\vspace{10pt}
\noindent\rule{\textwidth}{1pt}
\tableofcontents\thispagestyle{fancy}
\noindent\rule{\textwidth}{1pt}
\vspace{10pt}


\section{Introduction}

In statistical physics, we often encounter dynamic (or ``kinetic'') equations of the type
\begin{equation}
  \partial_t\phi = F(\phi,\nabla\phi,\dots;\vec x) + \eta
  \label{eq:phi:F}
\end{equation}
describing the stochastic evolution of a real-valued scalar field $\phi(\x,t)$ in $d$ dimensions ($\x\in\mathcal D\subseteq\mathbb R^d$). This could be the density change of a monocomponent fluid, related to the composition of a binary mixture, or the (relative) height of a film covering a surface. In all these cases, we might want to quantify the behavior at large scales. The details of the specific model under scrutiny are encoded in the function $F(\phi,\nabla\phi,\dots;\vec x)$ of the field and its derivatives, which we assume can be parametrized by a set of model parameters $\vec x=(x_1,\dots)$. Since $\phi(\x,t)$ is already a coarse-grained description of the microscopic degrees of freedom it is accompanied by a noise field $\eta(\x,t)$.

We will not discuss how to derive Eq.~\eqref{eq:phi:F} from a microscopic model, for which there is no unique answer (for an insightful discussion, see Ref.~\citenum{archer04}). Importantly, the functional form of $F$ is largely determined by symmetry considerations and conservation laws. This entails that rederiving the evolution equation on a coarser length scale, we obtain the same functional form $F(\phi';\vec x')$ but now with \emph{renormalized} parameters $\vec x'$, which thus become functions of the length scale, $\vec x\mapsto\vec x(\ell)$. The task of renormalization is to obtain the functional dependence for the model parameters. Mathematically, there is a clear analogy with the energy scale in particle physics, and indeed the confluence of both fields has led to one of the major breakthroughs, Wilson's renormalization group~\cite{wilson83}. Its success is due in no small part to the fact that fixed points of this renormalization ``flow'' encode the universal behavior of scale-free systems and close to critical points.

As an example, assuming that our model is invariant under the inversion $\phi\to-\phi$ as well as rotational symmetry, an expansion to lowest order yields
\begin{equation}
  F = -(\im\nabla)^\al(a\phi+u\phi^3-\kap\nabla^2\phi),
  \label{eq:F:gl}
\end{equation}
where we treat both cases of a non-conserved field ($\al=0$, known as model A) and a conserved field ($\al=2$, model B).\footnote{These names stem from the classification of Hohenberg and Halperin~\cite{hohenberg77}, see Table~I therein. A conserved quantity is one for which the integral $\Int{^d\x}\phi(\x,t)$ does not change with time. It implies that we have the continuity equation $\partial_t\phi+\nabla\cdot\vec j=0$ with current field $\vec j$.} This Ginzburg-Landau model is characterized by the parameters $\vec x=(a,u,\kap)$, the exact expressions of which follow from the specific microscopic model.

Here we describe how the renormalization procedure can be performed directly on the level of the evolution equation~\eqref{eq:phi:F}, which has been termed ``dynamic renormalization group''. A big advantage is that both equilibrium and (driven) non-equilibrium systems can be treated. Early applications include ferromagnets~\cite{ma75}, the stochastic Navier-Stokes and Burgers equation~\cite{forster77}, as well as the famous KPZ equation~\cite{kardar86}. More recent applications are on the conserved KPZ equation~\cite{caballero18a}, self-organized criticality~\cite{diaz-guilera94}, active nematics~\cite{mishra10}, and in particular on the collective behavior of self-propelled agents~\cite{toner95,toner98,gelimson15,caballero18,cavagna19,mahdisoltani21,speck22a}.

Actual calculations tend to look somewhat complicated and even intimidating. The purpose of these notes is to expose the minimal ``machinery'' to obtain one-loop flow equations and to study their properties. By no means are they a replacement for more detailed reviews~\cite{hohenberg77}, lecture notes~\cite{tauber12}, and books~\cite{goldenfeld92,tauber13} on renormalization. For completeness, we point out that stochastic dynamic equations can be transformed into a stochastic action (known as Doi-Peliti~\cite{doi76,peliti85} and Martin-Siggia-Rose~\cite{martin73,janssen76} formalisms), for which methods from (quantum) field theory are directly applicable. It does, however, add another layer of complexity and for the sake of simplicity we will not discuss it here.


\section{Setting the stage}

\subsection{Linear theory}
\label{sec:basics}

Neglecting all non-linear terms, we have $F_0=-(\im\nabla)^\al(a\phi-\kap\nabla^2\phi)$. We switch to Fourier space through
\begin{equation}
  \phi(\x,t) = \int\frac{\dd\om}{2\pi}\int\frac{\dd^d \vec q}{(2\pi)^d}e^{\im\vec q\cdot\x-\im\om t} \phi(\om,\vec q),
\end{equation}
where we use the same symbol for the field but with different arguments. The linearized evolution equation~\eqref{eq:phi:F} then reads
\begin{equation}
  -\im\om\phi = -q^\al(a+\kap q^2)\phi + \eta
  \label{eq:phi:lin}
\end{equation}
with solution $\phi(\om,\vec q)=G_0(\om,q)\eta(\om,\vec q)$, where we have defined the bare propagator
\begin{equation}
  G_0(\om,q) \equiv \frac{1}{-\im\om+q^\al(a+\kap q^2)} = \frac{1}{-\im\om+f(q)}
  \label{eq:G0}
\end{equation}
with $f(q)\equiv q^\al(a+\kap q^2)$ for later use. Throughout, we will write $q$ for the magnitude of the wave vector $\vec q$ (and analogously for other wave vectors). For the noise correlations, we will employ
\begin{equation}
  K(\hat q,\hat q') \equiv \mean{\eta(\hat q)\eta(\hat q')} \\ = 2Dq^\al(2\pi)^{d+1}\delta^d(\vec q+\vec q')\delta(\om+\om'),
  \label{eq:K}
\end{equation}
where, for ease of notation, we have combined frequency $\om$ and wave vector $\vec q$ into the single vector $\hat q\equiv(\om,\vec q)$ with $d+1$ components. Note that for a conserved field the factor $q^2$ suppresses fluctuations of the integrated field as required. The strength of the noise is quantified by the coefficient $D$. For dynamics obeying detailed balance, the fluctuation-dissipation theorem constraints $D$ to be related to the temperature but in the following we will mostly treat it as another free model parameter.

It is now straightforward to calculate the field correlations
\begin{equation}
  \mean{\phi(\hat q)\phi(\hat q')} = (2\pi)^{d+1}C_0(\hat q)\delta^{d+1}(\hat q+\hat q')
  \label{eq:corr}
\end{equation}
with the dynamic structure factor
\begin{equation}
  C_0(\om,q) \equiv 2Dq^\al G_0(\om,q)G_0(-\om,q) = \frac{2Dq^\al}{\om^2+[f(q)]^2}.
  \label{eq:C0}
\end{equation}
The static structure factor follows immediately as
\begin{equation}
  S_0(q) \equiv \int_{-\infty}^\infty\frac{\dd\om}{2\pi}C_0(\om,q) = \frac{D}{a+\kap q^2}
  \label{eq:S0}
\end{equation}
independent of $\al$. Clearly, we can construct one length scale, $\xi=(a/\kap)^{-1/2}$, which is the correlation length governing the exponential decay of correlations in real space. With this correlation length, the static structure factor becomes
\begin{equation}
  S_0(q) = \frac{(D/\kap)\xi^2}{1+(\xi q)^2}.
\end{equation}
For $a\to 0$ the correlation length $\xi\to\infty$ diverges with $S_0(q)=(D/\kap)q^{-2}$.

\subsection{Basic idea of renormalization}
\label{sec:rg}

Our model is useful down to a length scale $\Lam^{-1}$ (typically related to the particle size or the lattice spacing) below which we have no further information. In Fourier space this implies that $\phi(\om,q\geqslant\Lam)=0$. Let us collect the model parameters into the vector $\vec x=\vec x(\Lam)$ depending on the microscopic cut-off. Now we integrate out spatial features on length scales smaller than $b\Lam^{-1}$ with a factor $b>1$. We thus lose microscopic information (corresponding to large $q\sim\Lam$) and consequently will need new parameters $\vec x\mapsto\vec x'=\vec x(\Lam/b)$ to describe the evolution of the field.

\begin{figure}[t]
  \centering
  \includegraphics{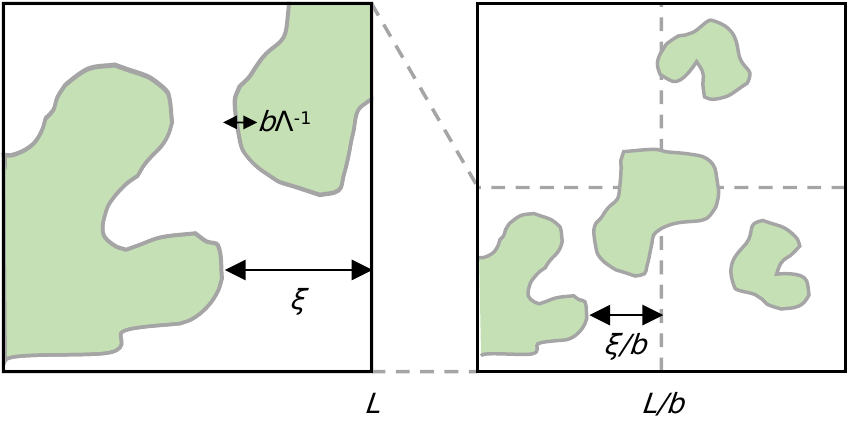}
  \caption{Sketch of the renormalization procedure in real space. (a)~We start with a system of length $L$, microscopic cut-off $\Lam^{-1}$, and correlation length $\xi$. After integrating out degrees of freedom on the smallest scales the cut-off length is increased to $b\Lam^{-1}$. (b)~We now zoom out by the factor $b$ to look at a larger portion of the system with restored cut-off $\Lam^{-1}$ and reduced correlation length $\xi/b$.}
  \label{fig:renorm}
\end{figure}

Before we repeat this step to get to an even coarser scale, let us ``zoom out'' by the factor $b$ and look at a larger portion of our system (Fig.~\ref{fig:renorm}). Holding the absolute size $L$ fixed, this means that coordinates $\x$ in our original system have shrunk to $\x/b$. Moreover, the size of any ``structure'' has shrunk by the same factor, in particular the correlation length $\xi\mapsto\xi'=\xi/b$. Note that this step restores the size of the smallest discernible features to $\Lam^{-1}$ (with respect to $L$).\footnote{As an analogy, consider looking at the system through a microscope at maximal magnification. You then reduce the magnification by a factor $b$ looking at a larger portion but with fixed field of view $L$ and fixed resolution $\Lam^{-1}$ (say, the size of a pixel). An excellent visualization for the Ising model can be found here: \url{https://www.youtube.com/watch?v=MxRddFrEnPc}.} Repeating this procedure induces an evolution, a ``flow'' in model space, during which we zoom out further and further. For an infinitesimal step $b=1+\delta\ell$ we find
\begin{equation}
  x'_i = x_i(\Lam/b) = x_i(\Lam) + \beta_i(\vec x)\delta\ell + \mathcal O(\delta\ell^2)
\end{equation}
with flow equations $\partial_\ell x_i=\beta_i(\vec x)$ implying the solution $\vec x(\ell)$ of model parameters as a function of $\ell$. Keeping explicitly track of the scale $b(\ell+\delta\ell)=b(\ell)(1+\delta\ell)$ yields $b=e^\ell$ and thus $\Lam(\ell)=\Lam_0e^{-\ell}$ is the actual value of the cut-off with initial value $\Lam_0$.

The remaining task is to find the functions $\beta_i$. Of particular interest are fixed points $\beta_i(\vec x^\ast)=0$ and the evolution around these fixed points. Their importance can be appreciated by noting that any initial correlation length $0<\xi_0<\infty$ will flow to $\xi(\ell)=\xi_0e^{-\ell}\to0$ except for points in our model space where the correlation length $\xi\to\infty$ diverges, which will be mapped to \emph{critical} fixed points.

\subsection{Scaling dimensions}
\label{sec:scale}

Implementing the rescaling step amounts to defining new wave vectors $q'=bq$ together with $\om'=b^z\om$, where $z$ is the dynamical exponent. Inserting both rescaled quantities into the bare propagator [Eq.~\eqref{eq:G0}] yields
\begin{equation}
  G_0(\om'/b^z,q'/b;\kap,a) = b^zG_0(\om',q';\kap'=b^{\Delta_\kap}\kap,a'=b^{\Delta_a}a)
\end{equation}
with $\Delta_\kap=z-\al-2$ and $\Delta_a=z-\al$. While the functional form of $G_0$ is the same, the model parameters have changed. Clearly, the correlation length $\xi'=(\kap'/a')^{1/2}=\xi/b$ indeed transforms as a length (with $\Delta_\xi=-1$). In the vicinity of critical points any quantity $x$ changes under rescaling as $x\mapsto b^{\Delta_x}x$ with $\Delta_x$ called the \emph{scaling dimension} of $x$. If a scaling dimension $\Delta_x<0$ is negative then it is called \emph{irrelevant}: Going to larger scales (large $b$) the influence of $x$ is diminished and eventually vanishes. Correspondingly, $\Delta_x>0$ is called \emph{relevant} and the borderline $\Delta_x=0$ \emph{marginal}. Demanding that the dynamic equation Eq.~\eqref{eq:phi:F} is invariant leads to another relation: the left side acquires a factor $b^{z+\Delta_\phi}$ and, therefore, the noise term scales with $b^{(\Delta_D+\al+d+z)/2}$ [cf. Eq.~\eqref{eq:K}] and thus $\Delta_D=2\Delta_\phi-d+z-\al$. Of particular importance is the Gaussian fixed point (G) at which \emph{all} non-linear terms become irrelevant and we are left with the linear theory introduced in Sec.~\ref{sec:basics}. 

To get a different view on scaling we briefly return to real space. The Fourier transform of $S_0(q)$ [Eq.~\eqref{eq:S0}] yields the static correlations
\begin{equation}
  \mean{\phi(\x)\phi(\x')} \sim \frac{1}{|\x-\x'|^{d-2}}
  \label{eq:phi:corr}
\end{equation}
for $|\x-\x'|\ll\xi$ ignoring numerical factors. If we demand that the functional form of these correlations does not change--we say that they are \emph{scale invariant}--then a rescaling $\x\mapsto\x/b$ of lengths on the right-hand side has to be compensated by a rescaling of the field, $\phi\mapsto b^{\Delta_\phi}\phi$, with ``naive'' scaling dimension $\Delta^0_\phi=(d-2)/2$. In general, the scaling dimension of the field is
\begin{equation}
  \Delta_\phi = \frac{d-2+\eta}{2}
  \label{eq:sd:phi}
\end{equation}
with $\eta$ known as the \emph{anomalous dimension} (not to be confused with the noise field). Moreover, assuming that $D$ is constant (as in thermal equilibrium, $\Delta_D=0$), we find $z=2+\al-\eta$~\cite{hohenberg77}.


\section{Graphical corrections}

\subsection{Perturbation series and vertex functions}

We now include non-linear terms into the evolution equation~\eqref{eq:phi:lin}. The $n$-th power of the field in Fourier space becomes
\begin{equation}
  [\phi(\x,t)]^n \to [\phi(\hat k_1)\cdots\phi(\hat k_n)]_{\hat q} = [\phi^n]_{\hat q} \\ = \left[\prod_{i=1}^n\int_{\hat k_i}\phi(\hat k_i)\right](2\pi)^{d+1}\delta(\hat q-\sum_{i=1}^n \hat k_i)
  \label{eq:phi:n}
\end{equation}
with integral
\begin{equation}
  \int_{\hat k} \equiv \int_{|\vec k|<\Lam}\frac{\dd\Om\dd^d \vec k}{(2\pi)^{d+1}}
\end{equation}
and $\hat k\equiv(\Om,\vec k)$.
Using this bracket notation, the evolution equation for the field reads
\begin{equation}
  \phi(\hat q) = G_0(\hat q)\eta(\hat q) + G_0(\hat q)\sum_{n=2,\dots}[v_n\phi^n]_{\hat q}
  \label{eq:phi}
\end{equation}
with vertex functions $v_n(\vec k_1,\dots,\vec k_n|\vec q)$. The dependence on wave vectors arises through spatial derivatives of the field in real space. The rule is to replace $\nabla\to-\im\vec q$ for $\nabla$'s acting on everything to their right and $\nabla\to\im\vec k_i$ inside brackets. For example, $\nabla^2\phi^2\to-q^2\phi(\hat k_1)\phi(\hat k_2)$ while $|\nabla\phi|^2\to-\vec k_1\cdot\vec k_2\phi(\hat k_1)\phi(\hat k_2)$. The vertex functions have to be symmetric with respect to exchanging wave vectors $\vec k_i$ since we have only one field.

Clearly, Eq.~\eqref{eq:phi} is not closed since it contains $\phi$ on both sides of the equation. Nevertheless, it can be used to generate the solution as a series of terms with increasing powers of the vertex strengths through inserting into itself. Assuming that these strengths are small implies a perturbation approach close to the Gaussian fixed point. The problem becomes clear immediately: relevant non-linear strengths grow under rescaling, taking them away from the region where the perturbative solution is valid. Our hope, thus, is to discover new fixed points in the vicinity of the Gaussian fixed point and to study their properties.

For consistency, it is helpful to define $v_1(q)\equiv G_0(0,q)=1/f(q)$ for the propagator and $v_0(q)\equiv C_0(0,q)$ for the correlation function. Here we have already set $\om=0$ since a Taylor expansion with respect to $\om$ simply generates derivatives with respect to $t$ in the time representation, which are absent in the original evolution equation~\eqref{eq:phi:F}. Inserting the solution Eq.~\eqref{eq:phi}, the goal is to determine how the vertex functions change after integrating out small-scale features, $[v_n\phi^n]_{\hat q}\to[\tilde v_n\phi^n]_{\hat q}$, with the change of $v_0\to\tilde v_0$ and $v_1\to\tilde v_1$ given through Eq.~\eqref{eq:corr} and Eq.~\eqref{eq:phi}, respectively.

\subsection{Constructing graphs}

\begin{figure}[t]
  \centering
  \includegraphics{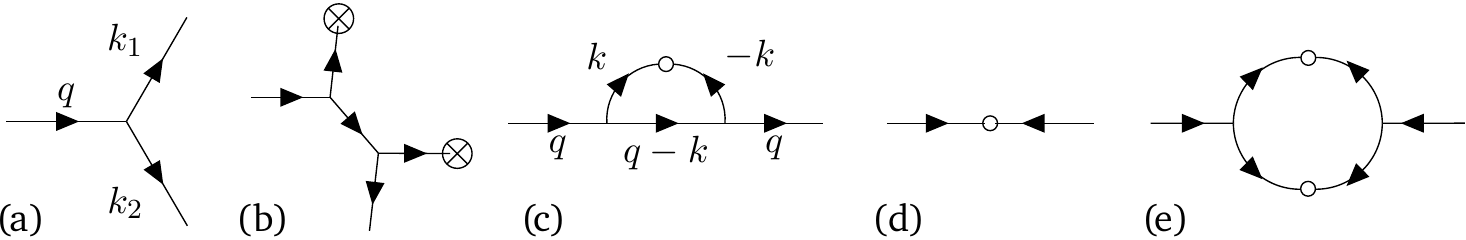}
  \caption{Intermediate steps to construct a graph. (a)~Initial bare vertex with $n=2$ fields as outgoing lines. (b)~Here we replace the upper line by the linear solution $\phi^+$ (indicated with a crossed dot) and attach a new vertex to the lower line. In the second step, we replace one of the two fields by $\phi^+$. (c)~We now join the two $\phi^+$ lines so that their wave vectors cancel, which completes this graph. The last step is to label all lines with their wave vectors. There are $2\times2=4$ ways to arrive at this graph. (d)~Correlation function $C_0$ and (e)~its first graphical correction due to two $2$-vertices.}
  \label{fig:graph}
\end{figure}

To keep track of the terms contributing to the solution it is helpful to employ a graphical language that is inspired by scalar Feynman diagrams (although it lacks the interpretation as particles and momenta)~\cite{weinzierl22}. Each term of the perturbation series is represented as a directed graph $\Gam_n$. We only need very few graphical elements: external lines, internal lines connecting vertices, and a ``sink'' of internal lines representing the correlation function $C_0$. Each vertex has one incoming line and $n$ outgoing lines [e.g., Fig.~\ref{fig:graph}(a) for $n=2$]. A single ($n=1$) outgoing line represents the propagator while the correlation function [Fig.~\ref{fig:graph}(d,e)] has only incoming lines ($n=0$). The sum of outgoing wave vectors $\sum_{i=1}^n\hat k_i=\hat q$ of each vertex equals the incoming wave vector $\hat q$, which is enforced by the $\delta$-distribution in Eq.~\eqref{eq:phi:n}.

To construct the final graph $\Gam_n$ from a bare initial vertex we iteratively either [Fig.~\ref{fig:graph}(b)]
\begin{itemize}
  \item replace a line by the linear solution $\phi\to\phi^+=G_0\eta$ or
  \item attach a vertex to one outgoing line (this line becomes the internal incoming line of the new vertex).
\end{itemize}
Finally, all intermediate $\phi^+$ lines need to end in an open dot, which joins exactly two lines so that the sum of their wave vectors vanishes [Fig.~\ref{fig:graph}(c)]. This step implicitly performs the average over the noise and the open dot plus the two lines together represent the correlation function $C_0(\hat k)$ [Fig.~\ref{fig:graph}(d)]. It should be easy to see that there are multiple ways to arrive at the same final graph $\Gam_n$. Section~\ref{sec:mult} shows how to calculate the multiplicity $|\Gam_n|$ as the number of permutations in the construction of the graph. For example, the multiplicity of the graph Fig.~\ref{fig:graph}(c) is $|\Gam_1|=2\times2=4$ because for each of the two steps there are two possibilities.

\subsection{From graph to integral}

The final graph can then be translated into one or several nested integrals $\mathcal I(\Gam_n;\vec x)$. All internal lines connecting two vertices represent $G_0(\hat k)$ with the corresponding wave vector. All external outgoing lines represent fields $\phi$ with one exception: If there is exactly one outgoing line ($n=1$) then its wave vector is necessarily $\hat q$ and it also represents $G_0(\hat q)$. In the following, for the final graphs we use the convention that the single incoming wave vector is $\vec q$, outgoing wave vectors are $\vec p_i$ with $\sum_i\vec p_i=\vec q$, and internal wave vectors are $\vec k_i$ that will be integrated out. Each internal wave vector necessarily implies a corresponding loop in the graph. For example, reading the graph Fig.~\ref{fig:graph}(c) from left to right leads to
\begin{equation}
  \mathcal I(\Gam_1;\vec x) = 4\int_{\hat k}
  G_0(\hat q)v_2(\vec k,\vec q-\vec k)C_0(\hat k)G_0(\hat q-\hat k)v_2(\vec q,-\vec k)G_0(\hat q)
  \label{eq:int:v2}
\end{equation}
and we need to integrate out $\hat k$ to obtain the lowest-order correction to the bare propagator $G_0(0,q)$. The pre-factor is the multiplicity of the graph.

Summing over all distinct graphs with the same number $n$ of outgoing lines yields the ``graphical corrections''
\begin{equation}
  \tilde v_n(\vec p_1,\dots,\vec p_n;\vec x,\Lam) = v_n(\vec p_1,\dots,\vec p_n;\vec x) + \sum_m\mathcal I(\Gam_n^{(m)};\vec p_1,\dots,\vec p_n;\vec x)
  \label{eq:graph_corr}
\end{equation}
for the vertex functions. We emphasize that the integrals $\mathcal I(\Gam_n)$ in general are functions of the outgoing wave vectors $\vec p_i$. To remain within the original model space (as defined by the function $F$), we have to reconstruct the functional form of the vertex $v_n$ neglecting terms involving higher orders of the wave vectors. The final step is to determine how the model parameters $x_i\to\tilde x_i$ change due to these graphical corrections by comparing the coefficients on both sides of Eq.~\eqref{eq:graph_corr}. While for simple vertex functions $v_n$ the $\tilde x_i$ can be read off, for functions that involve several outgoing wave vectors the problem can be cast as a system of linear equations (Sec.~\ref{sec:renorm_parameters}).

\subsection{Wilson's shell renormalization}

\begin{figure}[t]
  \centering
  \includegraphics{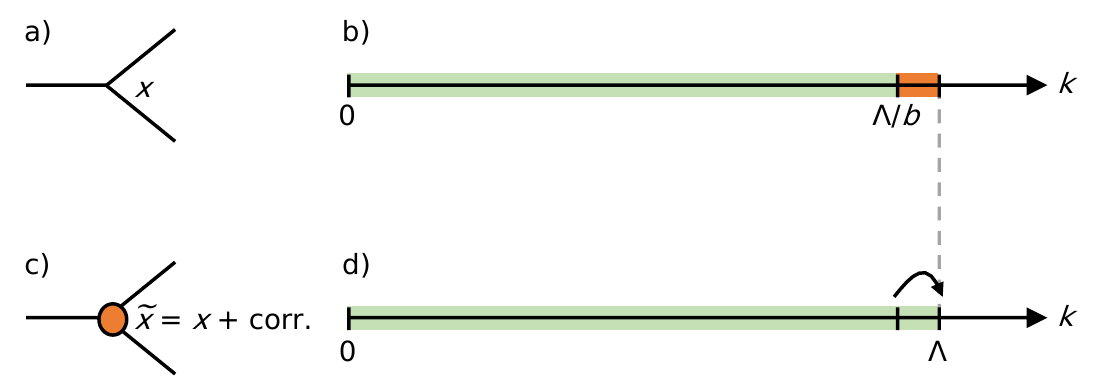}
  \caption{Sketch of Wilson's shell renormalization in Fourier space. (a)~A bare 2-vertex with coefficient $x$. (b)~We integrate out the wave vectors $k\in[\Lam/b,\Lam]$ corresponding to small length scales. (c)~This procedures a vertex with corrected coefficient $\tilde x$. (d)~All lengths are rescaled by the factor $b$ to restore the cut-off $\Lam$, cf. Fig.~\ref{fig:renorm}.}
  \label{fig:proc}
\end{figure}

The arguably most common scheme to implement the procedure sketched in Fig.~\ref{fig:renorm} is to consider an infinitesimal ``shell'' of wave vectors $k\in[\Lam/b,\Lam]$ through setting $b=1+\delta\ell$ and only consider contributions to linear order of $\delta\ell$. The first important consequence is that we only have to consider graphs with a single loop, and thus a single internal $\hat k$, since graphs with multiple loops are of higher order. Assuming that vertex functions are of the form $v_n=\sum_i x_iv_n^{(i)}$, from Eq.~\eqref{eq:graph_corr} we thus find $\tilde x_i=(1+\psi_{x_i}\delta\ell)x_i$ with $\psi_{x_i}(\vec x)$ quantifying the graphical one-loop corrections for $x_i$ due to the non-linearities. The next step is to restore the cut-off $\Lam/b\to\Lam$ through rescaling, implying
\begin{equation}
  x_i' = b^{\Delta_{x_i}}\tilde x_i = (1+\delta\ell)^{\Delta_{x_i}}(1+\psi_{x_i}\delta\ell)x_i \approx [1+(\Delta_{x_i}+\psi_{x_i})\delta\ell]x_i
\end{equation}
to linear order. Wilson's flow equations thus read
\begin{equation}
  \partial_\ell x_i = (\Delta_{x_i}+\psi_{x_i})x_i = \beta_i(\vec x).
  \label{eq:wilson:flow}
\end{equation}
Figure~\ref{fig:proc} summarizes the procedure.

\subsection{Handling the integrals}
\label{sec:integrals}

At this point we have to face the integral $\int_{\hat k}$. The first step is to integrate over the internal frequency $\Om$, which only involves the bare propagators. The generalization of Eq.~\eqref{eq:S0} for the product $C_0(\Om,k)\prod_iG_0(s_i\Om,k_i)$ of the bare propagators inside the loop reads
\begin{equation}
  \int_{-\infty}^\infty\frac{\dd\Om}{2\pi} \frac{2Dk^\al}{\Om^2+[f(k)]^2}\prod_{i=1}^p\frac{1}{-s_i\im\Om+f(k_i)} = \frac{Dk^\al}{f(k)}Q^{(p)}_{s_1\cdots s_n}\prod_{i=1}^p\frac{1}{f(k)+f(k_i)}
  \label{eq:I}
\end{equation}
with $\vec k_i$ the corresponding wave vector of the loop edge. Here, $s_i=\pm 1$ is the sign of $\hat k$ within $\hat k_i$ (remember that we set all external frequencies to zero). If all signs are equal then $Q^{(p)}=1$. For $p=2$ propagators and mixed signs one finds
\begin{equation}
  Q^{(2)}_{+-} = Q^{(2)}_{-+} = \frac{2f(k)+f(k_1)+f(k_2)}{f(k_1)+f(k_2)}
  \label{eq:Q}
\end{equation}
with similar but more complicated expressions for $p>2$ propagators.

The integral over the internal wave vector $\vec k$ is performed in spherical coordinates,
\begin{equation}
  \int\frac{\dd^d\vec k}{(2\pi)^d} = \frac{S_{d-1}}{(2\pi)^d}\IInt{k}{\Lam'}{\Lam}k^{d-1}\IInt{\theta}{0}{\pi}\sin^{d-2}\theta
  \label{eq:int:k}
\end{equation}
with polar angle $\theta$, $S_d\equiv2\pi^{d/2}/\Gam(d/2)$ the surface area of a unit hypersphere in $d$ dimensions, and $K_d\equiv S_d/(2\pi)^d$. Here, $\Gam(s)$ is the gamma function generalizing the factorial to non-integers. Useful angular integrals in the following are~\cite{gradshteyn2014table}
\begin{gather}
  \label{eq:S:0}
  S_{d-1}\IInt{\theta}{0}{\pi}\sin^{d-2}\theta = S_d, \\
  S_{d-1}\IInt{\theta}{0}{\pi}\sin^{d-2}\theta\cos\theta = 0, \\
  \label{eq:S:2}
  S_{d-1}\IInt{\theta}{0}{\pi}\sin^{d-2}\theta\cos^2\theta = \frac{S_d}{d}.
\end{gather}
Finally,
\begin{equation}
  \IInt{k}{\Lam/b}{\Lam}k^{d-1} h(k) = \Lam^d h(\Lam)\delta\ell + \mathcal{O}(\delta\ell^2)
  \label{eq:wilson:int}
\end{equation}
for any function $h(k)$ of the magnitude $k$.


\section{Illustrations}

\subsection{Model A/B: Wilson-Fisher fixed point}
\label{sec:modelB}

\begin{figure}[t]
  \centering
  \includegraphics{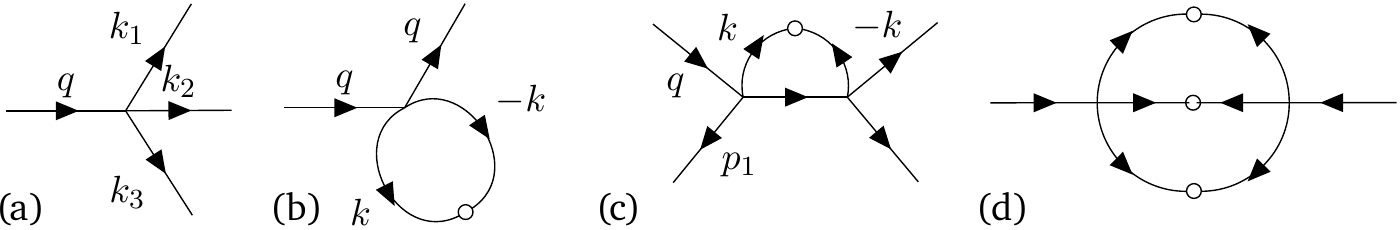}
  \caption{Relevant graphs of model A/B. (a)~Initial bare $3$-vertex. (b)~One-loop correction to the propagator. (c)~One-loop correction to the $3$-vertex involving two vertices. (d)~The first graphical correction to the noise strength is a two-loop integral.}
  \label{fig:modelB}
\end{figure}

To demonstrate how this machinery works in practice we first turn to model A/B [Eq.~\eqref{eq:F:gl}]. Let us see when $u$ becomes irrelevant, for which rescaling $a\phi+u\phi^3$ yields the scaling dimension $\Delta_u=\Delta_a-2\Delta_\phi$ and thus $\Delta^0_u=4-d$ with $\Delta^0_\phi=(d-2)/2$. In dimensions $d>4$ the non-linear term is irrelevant and the Gaussian fixed point is attractive. This changes for $d<4$ with $u(\ell)$ moving away from a small but non-zero initial $u_0$. We will now determine where it flows to. Model A/B has one non-zero vertex $v_3(q)=-uq^\al$ [Fig.~\ref{fig:modelB}(a)], which implies that graphs can only be constructed from $3$-vertices.

We start with the graph in Fig.~\ref{fig:modelB}(b), which contributes
\begin{equation}
  \mathcal I(\Gam_1) = -3uq^\al[G_0(\hat q)]^2\int_{\hat k}C_0(\hat k)
\end{equation}
to the propagator with multiplicity $|\Gam_1|=3$ since there are three ways to connect two out of three lines. The integral becomes
\begin{equation}
  \int_{\hat k}C_0(\hat k) = \int\frac{\dd^d\vec k}{(2\pi)^d}S_0(k) = K_d\Lam^dS_0(\Lam)\delta\ell
\end{equation}
with the static structure factor $S_0(k)$ given in Eq.~\eqref{eq:S0}. From $\tilde v_1(q)=v_1(q)+\mathcal I(\Gam_1)$ we find
\begin{equation}
  \tilde f(q) = f(q)\left[1 - 3uq^\al\frac{K_d\Lam^dS_0(\Lam)}{f(q)}\delta\ell\right]^{-1} \approx f(q) + 3uq^\al K_d\Lam^dS_0(\Lam)\delta\ell
\end{equation}
expanding again for small $\delta\ell\ll 1$. Plugging in $f(q)=q^\al(a+\kap q^2)$, we can now read off the graphical corrections of the model parameters with intermediates $\tilde\kap=\kap$ (whence $\psi_\kap=0$) and
\begin{equation}
  \tilde a = a + 3u K_d\Lam^dS_0(\Lam)\delta\ell, \qquad \psi_a = \frac{3uD}{a}\frac{K_d\Lam^d}{a+\kap\Lam^2}
\end{equation}
due to Fig.~\ref{fig:modelB}(b).

We now turn to the graph in Fig.~\ref{fig:modelB}(c). The multiplicity of this graph is: $|\Gam_3|=3$ (possibilities to insert the new vertex) $\times 2$ (remaining possibilities to insert a noise) $\times 3$ (possibilities to insert a noise in the new vertex). Reading from left to right vertex, we have
\begin{equation}
  \mathcal I(\Gam_3) = 18v_3(q)\int_{\hat k}C_0(\hat k)G_0(\hat q-\hat k-\hat p_1)v_3(\vec q-\vec k-\vec p_1)
  \label{eq:int:v3}
\end{equation}
leaving out the external lines (they are not part of the function $v_3$). We can immediately set $\hat p_1\to0$ inside the integral since also $\tilde v_3(q)=-\tilde uq^\al$ will only depend on $q$. The frequency integral of $C_0(\hat k)G_0(\hat q-\hat k)$ then reads [cf. Eq.~\eqref{eq:I}]
\begin{equation}
  \int_{-\infty}^\infty\frac{\dd\Om}{2\pi}\frac{2Dk^\al}{\Om^2+[f(k)]^2}\frac{1}{\im\Om+f(\vec q-\vec k)} = \frac{Dk^\al}{f(k)[f(k)+f(\vec q-\vec k)]}.
  \label{eq:int:CG}
\end{equation}
Since the pre-factor in Eq.~\eqref{eq:int:v3} is already $\propto q^\al$ we can let $q\to0$ for the remaining terms inside the integral, which yields $\tilde v_3(q)=v_3(q)+\mathcal I(\Gam_3)$ and thus
\begin{equation}
  \tilde u = u - 9u^2D\frac{K_d\Lam^d}{(a+\kap\Lam^2)^2}\delta\ell, \qquad \psi_u = -9uD\frac{K_d\Lam^d}{(a+\kap\Lam^2)^2}.
\end{equation}
A quick look at Fig.~\ref{fig:modelB}(d) reveals that the first correction to $D$ is already a two-loop integral and thus of order $(\delta\ell)^2$ with $\psi_D=0$.

\begin{figure}[t]
  \centering
  \includegraphics{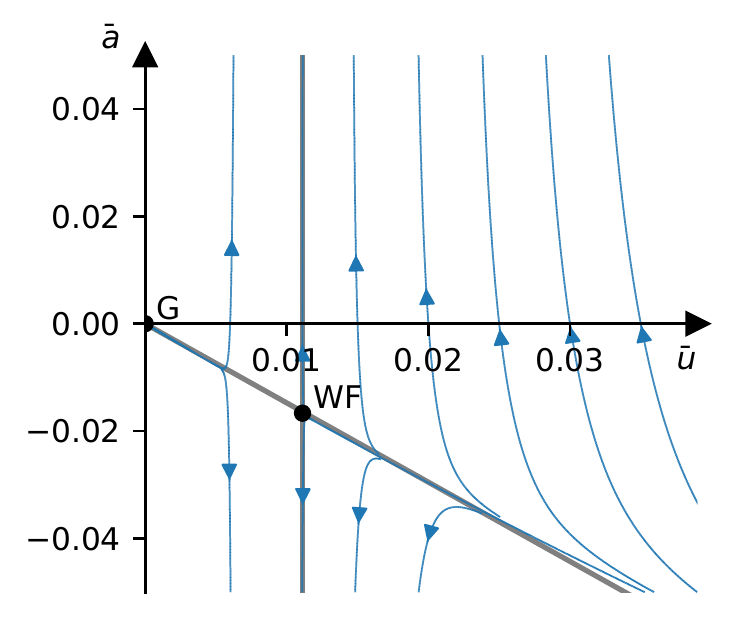}
  \caption{Streamline plot of the flow equations~\eqref{eq:flow} for model A/B using $\eps=0.1$. Indicated are the Gaussian fixed point (G) and the Wilson-Fisher fixed point (WF) with the eigenvectors along the gray lines.}
  \label{fig:flow}
\end{figure}

For the final flow equations we introduce the reduced ``dimensionless'' model parameters
\begin{equation}
  \bar a \equiv \frac{a}{\kap\Lam^2}, \qquad
  \bar u \equiv \frac{uD}{\kap^2}K_d\Lam^{-\eps}
\end{equation}
leading to
\begin{equation}
  \partial_\ell\bar u = \left(\frac{\partial_\ell u}{u}+\frac{\partial_\ell D}{D}-2\frac{\partial_\ell\kap}{\kap}\right)\bar u = (\Delta_u+\Delta_D-2\Delta_\kap+\psi_u)\bar u
\end{equation}
through inserting their flow equations~\eqref{eq:wilson:flow}. Note how this choice removes the unknown exponents since with the scaling relations derived in Sec.~\ref{sec:scale} we find $\Delta_a-\Delta_\kap=2$ and $\Delta_u+\Delta_D-2\Delta_\kap=4-d=\eps$. At this point the dimension $d$ is simply a number and not necessarily an integer. This freedom is exploited to introduce the small parameter $\eps$ with non-integer dimension $d=4-\eps$ close to the upper critical dimension at which the Gaussian fixed point becomes repulsive. The final flow equations then read
\begin{equation}
  \partial_\ell\bar a = 2\bar a + 3\bar u\frac{1}{\bar a+1}, \qquad
  \partial_\ell\bar u = \left(\eps - 9\bar u\frac{1}{(\bar a+1)^2}\right)\bar u
  \label{eq:flow}
\end{equation}
expressed in the reduced parameters. Besides the Gaussian fixed point ($\bar a=\bar u=0$) these equations admit another fixed point of order $\eps$, the Wilson-Fisher (WF) fixed point located at $\bar u^\ast=\eps/9$ and $\bar a^\ast=-\eps/6$ to linear order~\cite{wilson72}. The resulting flow of the reduced parameters is plotted in Fig.~\ref{fig:flow}. For this model, the perturbative approach thus has been successful since the non-linear parameter $u\sim\eps$ remains small and does not run off.

To understand the flow around WF, we define $\tau\equiv\bar a-\bar a^\ast$ and $\delta\bar u=\bar u-\bar u^\ast$ measuring the distance to the critical point with new model parameters $\vec x=(\tau,\delta\bar u)^T$. The coupled linearized flow equations read
\begin{equation}
  \partial_\ell\left(\begin{array}{c}
    \tau \\ \delta\bar u    
  \end{array}\right) = \left(\begin{array}{cc}
    2-\tfrac{\eps}{3} & 3(1+\tfrac{\eps}{6}) \\ 0 & -\eps
  \end{array}\right)
  \left(\begin{array}{c}
    \tau \\ \delta\bar u    
  \end{array}\right)
\end{equation}
up to linear order of $\eps$. This matrix has two eigenvalues. The first is $\Delta_{\delta u}=-\eps$ and thus becomes irrelevant for $d<4$. As a consequence, the flow of $u$ is repelled from G and flows towards WF along its eigenvector. The second eigenvector is $\vec x_\tau=(1,0)^T$ so we identify its eigenvalue with $\Delta_\tau=2-\eps/3$. We expect the correlation length to diverge as $\xi\sim|\tau|^{-\nu}$ with exponent $\nu$. From the scaling dimensions (Sec.~\ref{sec:scale}) we immediately find $\Delta_\xi=-\nu\Delta_\tau=-1$ and thus $\nu=1/\Delta_\tau\approx\tfrac{1}{2}+\tfrac{\eps}{12}$. While strictly valid only close to $d=4$, the expansion in $\eps$ can be improved systematically and there is ample evidence that the WF controls the critical behavior down to $d=2$~\cite{leguillou85}.

\subsection{Burgers-KPZ equation}

In a seminal paper Kardar, Parisi, and Zhang (KPZ) proposed that the equation
\begin{equation}
  \partial_t\phi = \kap\nabla^2\phi + \frac{\lam}{2}|\nabla\phi|^2 + \eta
  \label{eq:phi:kpz}
\end{equation}
governs the non-equilibrium dynamics of a coarse-grained height field $\phi(\x,t)$ above a $d$-dimensional substrate~\cite{kardar86,takeuchi18}. In this context, $\kap$ is the surface tension smoothing the surface and $\lam$ determines the local growth rate along the surface normal. The field is not conserved and thus $\al=0$. Moreover, $a=0$, implying scale-free growth with $-\Delta_\phi$ the roughness exponent determining whether the surface is smooth ($\Delta_\phi>0$) or rough ($\Delta_\phi<0$).

Invariance of Eq.~\eqref{eq:phi:kpz} under scaling yields the scaling relations
\begin{equation}
  \Delta_\kap = z-2, \qquad \Delta_\lam = z-2-\Delta_\phi, \qquad \Delta_D = 2\Delta_\phi - d + z.
\end{equation}
Importantly, Eq.~\eqref{eq:phi:kpz} enjoys Galilean invariance through replacing $\phi\to\phi+\bm\epsilon\cdot\x$ together with $\x\to\x+\bm\epsilon\lam t$ describing an infinitesimal tilting of the surface. This symmetry implies that the coefficient $\lam$ cannot receive graphical corrections, $\psi_\lam=0$~\cite{medina89}.

Let us calculate the remaining corrections $\psi_i$. The only non-zero vertex is given through $v_2(\vec k_1,\vec k_2)=-(\lam/2)\vec k_1\cdot\vec k_2$. For the graph shown in Fig.~\ref{fig:graph}(c) we have already calculated the frequency integral in Eq.~\eqref{eq:int:CG}. Plugging this result with $f(k)=\kap k^2$ into Eq.~\eqref{eq:int:v2} leads to
\begin{equation}
  \mathcal I(\Gam_1) = -\frac{\lam^2D}{\kap^2}[G_0(\hat q)]^2\int\frac{\dd^d\vec k}{(2\pi)^d}\frac{\vec k\cdot(\vec q-\vec k)(\vec q\cdot\vec k)}{k^2(k^2+|\vec q-\vec k|^2)}.
  \label{eq:kpz:int}
\end{equation}
To calculate this integral, we symmetrize $\vec k\to\vec k+\vec q/2$ and expand in powers of $x=q/k$ with $\vec q\cdot\vec k=k^2x\cos\theta$. To lowest order the integrand reads
\begin{equation}
  \frac{(\tfrac{x^2}{4}-1)(\tfrac{x^2}{2}+x\cos\theta)}{2\kap^2(1+\tfrac{x^2}{4}+x\cos\theta)(1+\tfrac{x^2}{4})} \approx \frac{x^2}{2\kap^2}(\cos^2\theta-\tfrac{1}{2}).
\end{equation}
Performing the angular integrals [Eqs.~\eqref{eq:S:0} and~\eqref{eq:S:2}] and reading off $\tilde\kap$ leads to
\begin{equation}
  \psi_\kap = \frac{\lam^2D}{\kap^3}K_d\Lam^{d-2}\frac{2-d}{4d} = \bar\lam^2\frac{2-d}{4d}.
  \label{eq:kpz:kap}
\end{equation}
Importantly, since the lowest order is $q^2$ this graph only corrects $\kap$ but does not generate a correction to $a$, which remains zero. Here we have defined the sole dimensionless parameter $\bar\lam^2\equiv\lam^2D\kap^{-3}K_d\Lam^{d-2}$.

Turning to the graph~\ref{fig:graph}(e), now the noise does receive corrections through
\begin{equation}
  \mathcal I(\Gam_0) = 2G_0(\hat q)G_0(-\hat q)\int_{\hat k} [v_2(\vec k,-\vec k)C_0(\hat k)]^2 = 2G_0(\hat q)G_0(-\hat q)\frac{\lam^2D^2}{4\kap^3}K_d\Lam^{d-2}\delta\ell.
\end{equation}
Here something new has happened since the graph includes two correlation functions. The possible permutations of noise terms is described by Isserlis' (or Wick's) theorem and yields the additional multiplicity factor two. Comparing with the expression for $C_0(\hat q)$ [Eq.~\eqref{eq:C0}], we read off the correction
\begin{equation}
  \psi_D = \frac{\lam^2D}{4\kap^3}K_d\Lam^{d-2} = \frac{\bar\lam^2}{4}
  \label{eq:kpz:D}
\end{equation}
for the noise strength $D$. The flow equation for our single dimensionless model parameter reads
\begin{equation}
  \partial_\ell\bar\lam = \left(\frac{\partial_\ell \lam}{\lam}+\frac{1}{2}\frac{\partial_\ell D}{D}-\frac{3}{2}\frac{\partial_\ell\kap}{\kap}\right)\bar\lam = \left(\Delta_\lam + \frac{1}{2}\Delta_D - \frac{3}{2}\Delta_\kap + \frac{1}{2}\psi_D - \frac{3}{2}\psi_\kap\right)\bar\lam.
\end{equation}
Plugging in the scaling relations for the $\Delta_i$ and the graphical corrections [Eqs.~\eqref{eq:kpz:kap} and~\eqref{eq:kpz:D}] leads to the closed flow equation~\cite{kardar86}
\begin{equation}
  \partial_\ell\bar\lam = \left(\frac{2-d}{2}+\bar\lam^2\frac{2d-3}{4d}\right)\bar\lam
\end{equation}
for the non-linear coefficient. We recover the Gaussian fixed point at $\bar\lam=0$, which is attractive for $d>2$ and repulsive for $d<2$. In $d>2$ dimensions there is a second perturbative fixed point at $|\bar\lam^\ast|=\sqrt{-4\eps}$ to lowest order in $\eps=2-d$, which is connected to a dynamic phase transition (``roughening transition''). For a discussion, we refer to the literature, e.g. Ref.~\cite{frey94}.

\subsection{A neural network model}

\begin{figure}[t]
  \centering
  \includegraphics{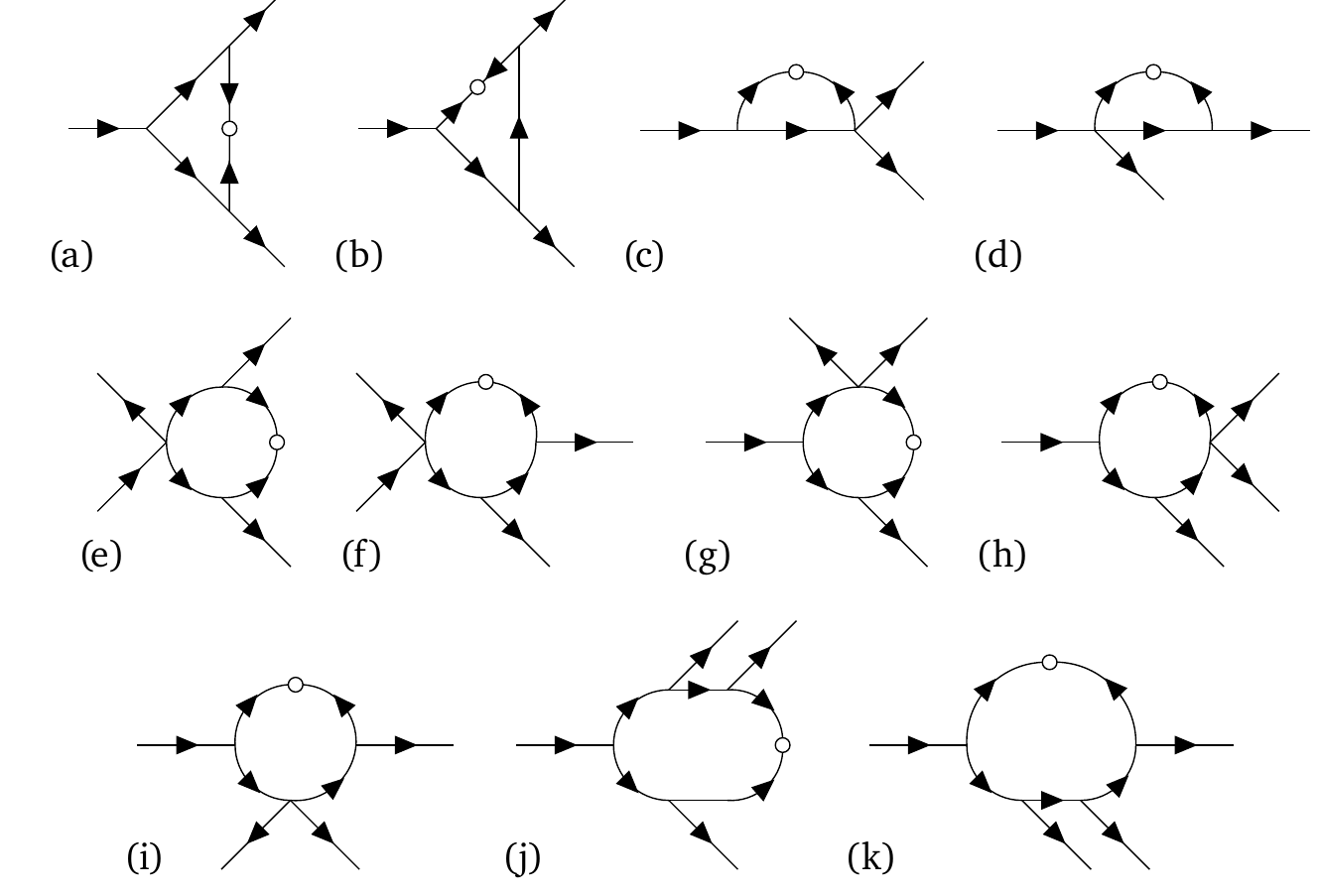}
  \caption{Additional one-loop graphs due to non-vanishing $v_2$ and $v_3$. (a-d)~Corrections contributing to $\tilde v_2$. (e-k)~Corrections contributing to $\tilde v_3$.}
  \label{fig:mixed}
\end{figure}

As the third and final example, we consider
\begin{equation}
  \partial_t\phi = \nabla^2(\kap\phi + c_2\phi^2 + c_3\phi^3) + \eta
\end{equation}
with non-conserved noise ($\al=0$). This evolution equation has been derived recently in Ref.~\citenum{tiberi22} for the evolution of a neural network (related to the famous Wilson-Cowan model~\cite{wilson72a}), where $\phi(\x,t)$ is a neural activity field. We have $f(q)=\kap q^2$, $v_2(q)=-c_2q^2$, and $v_3(q)=-c_3q^2$. Hence, we now have to consider all graphs that can be constructed from both $2$-vertices and $3$-vertices, which are depicted in Fig.~\ref{fig:mixed}.

All corrections are at order $q^2$, which immediately implies $\psi_D=0$ and we can set all external wave vectors to zero within the loop integral. For the propagator, the graph Fig.~\ref{fig:graph}(c) contributes
\begin{equation}
  \mathcal I(\Gam_1^{(2)}) = 4[G_0(\hat q)]^2\int\frac{\dd^d\vec k}{(2\pi)^d}\frac{Dv_2(q)v_2(\vec q-\vec k)}{\kap k^2(\kap k^2+\kap |\vec q-\vec k|^2)} = \frac{2c_2^2D}{\kap^2}[G_0(0,q)]^2q^2K_d\Lam^{d-2}\delta\ell
\end{equation}
and together with Fig.~\ref{fig:modelB}(b) we obtain $\psi_\kap=3\bar c_3-2\bar c_2^2$. As before, we define dimensionless parameters
\begin{equation}
  \bar c_2^2 \equiv \frac{c_2^2D}{\kap^3}K_d\Lam^{d-2}, \qquad \bar c_3 \equiv \frac{c_3D}{\kap^2}K_d\Lam^{d-2}.
\end{equation}
Similar to the KPZ model, no correction $\psi_a=0$ is generated and $a$ remains zero with diverging correlation length throughout.

Let us move on to $\tilde v_2$. The integral of Fig.~\ref{fig:mixed}(a) is
\begin{equation}
  \mathcal I(\Gam_2^{(1)}) = 4\int\frac{\dd^d\vec k}{(2\pi)^d}\frac{DQ^{(2)}_{+-}v_2(q)v_2(k)v_2(k)}{\kap k^2(2\kap k^2)^2} = -\frac{2c_2^3D}{\kap^3}q^2K_d\Lam^{d-2}\delta\ell
\end{equation}
with $Q^{(2)}_{+-}=2$ from Eq.~\eqref{eq:Q}. Since we set all external wave vectors to zero, $Q^{(p)}$ reduces to a number. For $p$ propagators, the denominator contributes a factor $2^p$ so that the prefactor of all integrals is simply $|\Gam|Q^{(p)}/2^p$ and the integrals are only distinguished by the products of $c_2$ and $c_3$. In Table~\ref{tab:graphs}, we summarize all graphs together with their multiplicity and the signs. From this table, we can essentially read off the integration result for all further graphs, leading to\footnote{Note that for the second and third term of $\psi_3$ we find slightly different coefficients from Ref.~\cite{tiberi22}.}
\begin{equation}
  \psi_2 = 4\bar c_2^2 - 9\bar c_3, \qquad
  \psi_3 = -9\bar c_3 + 21\bar c_2^2 - 5\bar c_2^4/\bar c_3.
\end{equation}
The final flow equations read
\begin{gather}
  \label{eq:flow:c2}
  \partial_\ell\bar c_2^2 = (2\Delta_2+\Delta_D-3\Delta_\kap + 2\psi_2-3\psi_\kap)\bar c_2^2 = (2-d+14\bar c_2^2-27\bar c_3)\bar c_2^2 \\
  \label{eq:flow:c3}
  \partial_\ell\bar c_3 = (\Delta_3+\Delta_D-2\Delta_\kap + \psi_3-2\psi_\kap)\bar c_3 = (2-d - 15\bar c_3+25\bar c_2^2)\bar c_3 - 5\bar c_2^4
\end{gather}
for the two dimensionless non-linear parameters.

\begin{table}
  \centering
  \begin{tabular}{cccccc}
    & figure & multiplicity & order & signs \\
    \hline
    $\tilde v_0$ & 2(e) & 2 & $v_2^2$ \\
    \hline
    $\tilde v_1$ & 2(c) & 4 & $v_2^2$ \\
                 & 4(b) & 3 & $v_3$ \\
    \hline
    $\tilde v_2$ & 6(a) & 4 & $v_2^3$ & $+-$ \\
                 & 6(b) & 8 & $v_2^3$ & $++$ \\
                 & 6(c) & 6 & $v_3v_2$ \\
                 & 6(d) & 12 & $v_3v_2$ \\
    \hline
    $\tilde v_3$ & 4(c) & 18 & $v_3^2$ \\
                 & 6(e) & 12 & $v_3v_2^2$ & $+-$ \\
                 & 6(f) & 24 & $v_3v_2^2$ & $++$ \\
                 & 6(g) & 6 & $v_3v_2^2$ & $+-$ \\
                 & 6(h) & 12 & $v_3v_2^2$ & $++$\\
                 & 6(i) & 12 & $v_3v_2^2$ & $++$\\
                 & 6(j) & 8 & $v_2^4$ & $++-$ \\
                 & 6(k) & 16 & $v_2^4$ & $+++$\\
  \end{tabular}
  \caption{Summary of the graphical corrections for all distinct one-loop graphs that can be constructed from 2-vertices and 3-vertices. The multiplicity is for symmetric graphs with indistinguishable outgoing lines (no dependence on external wave vectors). The last column specifies the signs of the integration variable $\hat k$ if the loop consists of more than one propagator, cf. Sec.~\ref{sec:integrals}.}
  \label{tab:graphs}
\end{table}

\begin{figure}[t]
  \centering
  \includegraphics[width=1\textwidth]{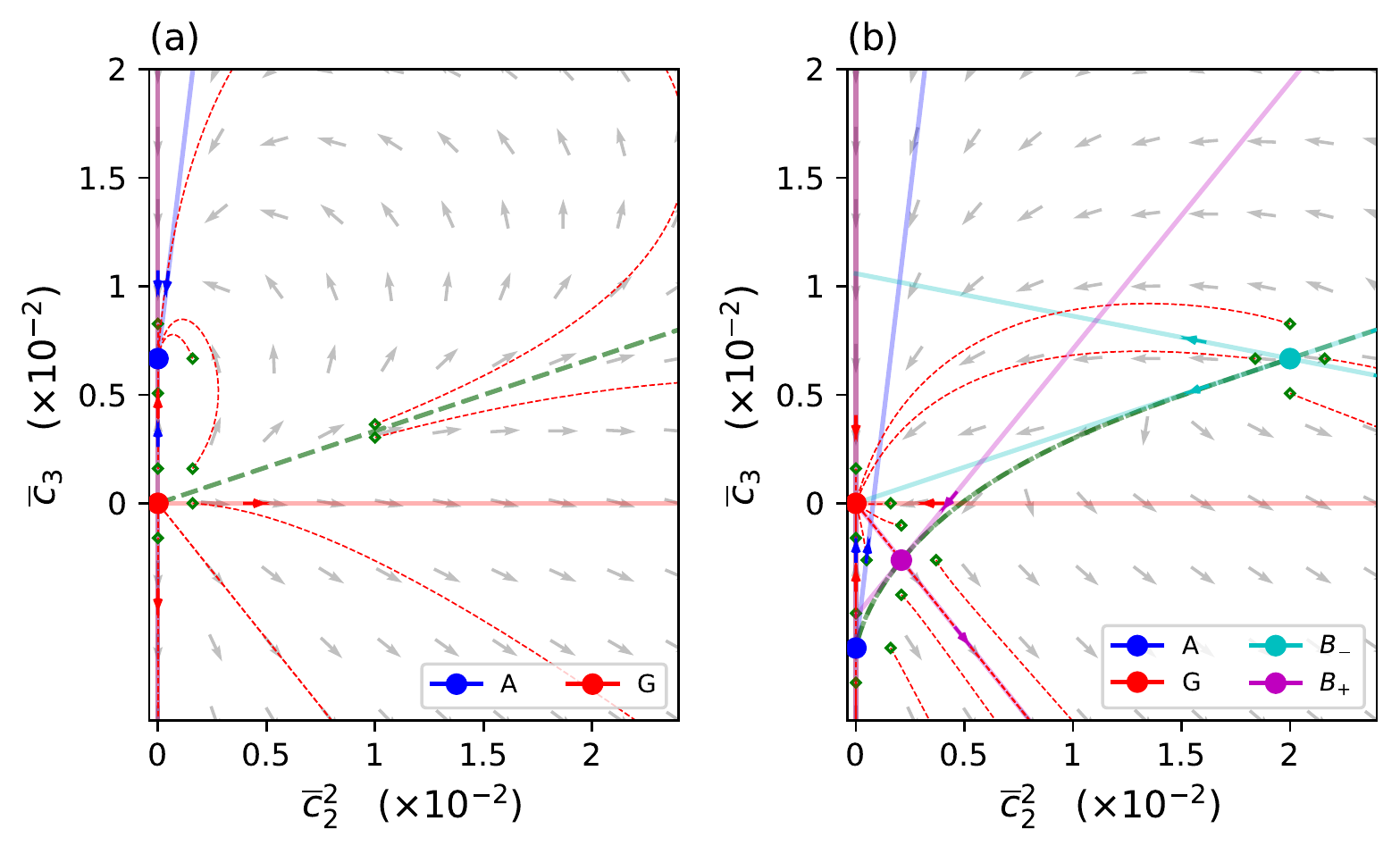}
  \caption{Streamline plot of the flow equations \eqref{eq:flow:c2} and \eqref{eq:flow:c3} for (a)~$\eps=0.1$ ($d<2$) and (b)~$\eps=-0.1$ ($d>2$). Closed symbols indicate the corresponding fixed points and the green open symbols indicate the initial points of a few trajectories (dotted lines). The eigenvectors of the fixed points are the straight lines denoted by the same color as the fixed point. While $B_-$ is repulsive, $B_+$ has one attractive and one repulsive direction. The separatrix is shown as the green dashed line: trajectories starting above the separatrix flow into the attractive fixed point ($A$ for $d<2$ and $G$ for $d>2$) while trajectories starting below run off to infinity.}
  \label{fig:neural}
\end{figure}

In Fig.~\ref{fig:neural}, we plot the flow for two values of $\eps=2-d$ in the plane $(\bar c_2^2,\bar c_3)$. The Gaussian fixed point $G$ at $(0,0)$ is attractive for $d>2$ and becomes repulsive for $d<2$. For $\bar c_2^2=0$ we find another perturbative fixed point $A$ at $(0,\eps/15)$, which is attractive for $d<2$ and now governs the flow in the vicinity of $G$. For $d>2$, $A$ becomes repulsive and we now find two more perturbative fixed points $B_\pm$ from the quadratic equation [after eliminating $\bar c_2^2$ by setting Eq.~\eqref{eq:flow:c2} to zero]
\begin{equation}
  \frac{2865}{196}\bar c_3^2 + \frac{29}{49}\eps\bar c_3 - \frac{5}{196}\eps^2 = 0
\end{equation}
with solutions $\bar c_3^\ast=-\eps/15$ and $\bar c_3^\ast=5\eps/191$ and thus $B_+$ at $(-0.021\eps,0.026\eps)$ and $B_-$ at $(-0.200\eps,-0.067\eps)$. These fixed points require $\eps\leqslant 0$ so that $\bar c_2^2\geqslant 0$.

Starting from an initial point in the plane, there are two possible behaviors: either the flow runs into an attractive fixed point ($A$ for $d<2$ and $G$ for $d>2$) or the flow runs off to infinity. This could be an artifact of the one-loop approximation (Ref.~\cite{tiberi22} finds numerical evidence for $d=2$ that this is indeed the case) or it indicates the existence of a strong-coupling fixed point that is not accessible in our perturbative approach. Both behaviors are delineated by a line, the \emph{separatrix}.

To calculate the separatrices, we recast Eqs.~\eqref{eq:flow:c2} and \eqref{eq:flow:c3} into a single differential equation $\frac{d \bar c_2^2}{d \bar c_3}$, which can be solved through an appropriate change of variables and substitution. For $\epsilon>0$ it can be shown that the line $y=\frac{1}{3}x$ is the separatrix. For $\epsilon<0$, the separatrix was found by integrating the flow equations backward in time starting close to the saddle point $B_+$ and using linear analysis close to the unstable points $B_-$ and $A$. While a comprehensive discussion of the implications is beyond the scope of these notes, this example demonstrates how fixed points shape the possible large-scale behavior.


\section{Conclusions}

We hope that we could show that calculating the one-loop flow equations for coarse-grained continuum models is conceptually not too difficult. Instead of focusing on (numerically) solving the evolution equations for a fixed set of parameters, these flow equations give insight into the possible large-scale behavior. Here we have restricted our attention to models up to order $\phi^3$ of the scalar field implying $2$-vertices and $3$-vertices, which already encompasses a large number of relevant models. Out of these we have presented three specific models, all of which have rather simple vertex functions. Going beyond and working with complex vertex functions that depend on several outgoing wave vectors is tedious and error prone, but can be tackled efficiently using symbolic computer algebra systems. A recent example is the extension of model B to active systems~\cite{caballero18,speck22a}. As a companion to these notes, we have developed the package ``REnormalization in STatistical physics and FLOW equations'' (\emph{restflow})\footnote{You can access the code at \url{https://github.com/us-itp4/restflow}.} implemented in python and based on sympy~\cite{meurer17} that implements the techniques discussed here.

While numerical investigations of the macroscopic phase behavior in complex systems are indispensable, the flow equations yield complementary \emph{qualitative} insights. Moreover, for driven systems governed by dynamics breaking detailed balance, many of the advanced numerical methods to improve sampling are not available and one has to resort to ``vanilla'' molecular dynamics. In this case renormalization offers a compelling alternative over computationally expensive simulations. The different phases are governed by the topology of fixed points in the space spanned by the model parameters. Critical fixed points with a diverging correlation length are of particular interest. In model A/B, we had to tune $a$ (typically related to the temperature) to reach the fixed point, which implies that it indeed corresponds to a critical point in the phase diagram. The other two models exhibit scale-free behavior throughout their parameter space.

Truncating at one-loop can introduce artifacts, e.g. the absence of a finite fixed point in the KPZ model for $3/2<d<2$~\cite{frey94}. A compelling perspective is to adapt recent advances in the calculation of scalar Feynman graphs~\cite{weinzierl22}, such as master integrals and using that Feynman integrals obey differential equations~\cite{henn13}, to statistical physics models and to move beyond one-loop graphs.

\appendix

\section{Calculation of multiplicity for one-loop graphs}
\label{sec:mult}

\begin{figure}[t]
  \centering
  \includegraphics{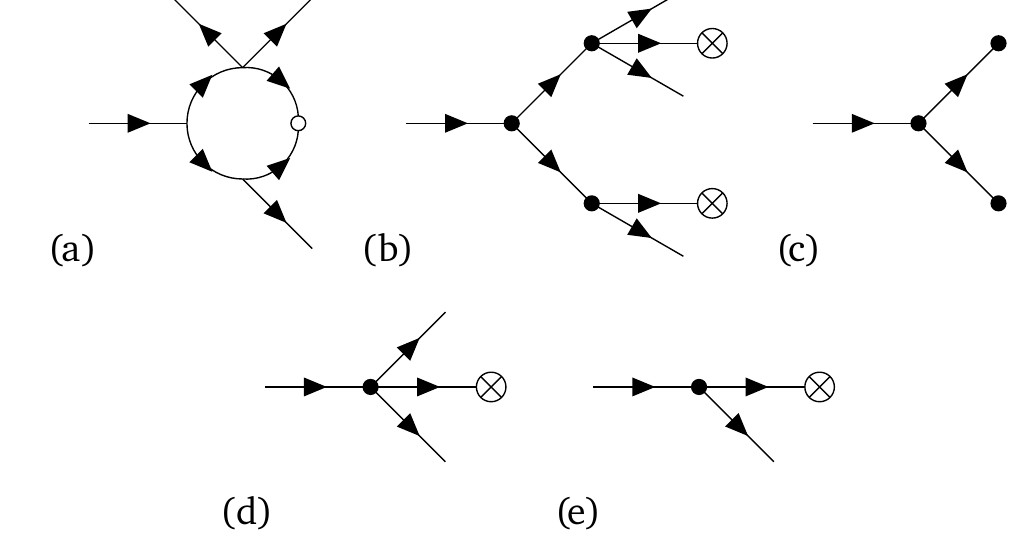}
  \caption{Example to illustrate the steps to calculate the multiplicity. (a)~Original graph [same as Fig.~\ref{fig:mixed}(g)]. (b)~The corresponding tree after cutting the open dot. (c-e)~The three vertices.}
  \label{fig:mult}
\end{figure}

To calculate the multiplicity of a given one-loop graph $\Gam$, we cut the correlation function and arrange the vertices into a tree. Each edge ends in either: another vertex (black dot), a noise term (crossed dot), or nothing. For each vertex $v_i$, the number of possible permutations of these three symbols is
\begin{equation}
  N_i = \frac{E!}{B!C!(E-B-C)!}
  \label{eq:perm}
\end{equation}
with $E$ the number of outgoing edges, $B$ the number of black dots, and $C$ the number of crossed dots. The graph multiplicity is $|\Gam|=\prod_i N_i$.

As an example, let us consider the graph Fig.~\ref{fig:mult}(a), which has three vertices in total. Its tree is shown in Fig.~\ref{fig:mult}(b). In Fig.~\ref{fig:mult}(c-e), we have isolated the vertices and their edges. Fig.~\ref{fig:mult}(c) has $E=2$, $B=2$, and $C=0$ with $N_1=1$. Fig.~\ref{fig:mult}(d) has $E=3$, $C=1$, and $B=0$ with $N_2=3$. Fig.~\ref{fig:mult}(e) has $E=2$, $C=1$, and $B=0$ with $N_3=2$. Multiplying the number of permutations, we obtain $|\Gam|=6$.

\section{Reconstructing vertex functions}
\label{sec:renorm_parameters}

Given the model parameters $\vec x$, let us assume that we have calculated the sum $I(\vec p_1,\dots,\vec p_n;\vec x)=\sum_m\mathcal I(\Gam^{m})$ of the one-loop graph integrals. We focus on a 2-vertex $v_2(\vec q,\vec p;\vec x)$ with scalar product $\vec q\cdot\vec p=qp\cos\psi$ but the following scheme extends to higher vertices. As already mentioned, we assume that $v_2=\sum_i x_iv_2^{(i)}$ is linear in the model parameters. Since vertex functions are polynomials, we expand $v_2^{(i)}=\sum_k a^{(i)}_kf_k$ in the monomial basis $f_k=\{q^2,p^2,qp\cos\psi,\dots\}$ with constant coefficients $a^{(i)}_k$. We expand the integral $I=\sum_kb_k(\vec x)f_k\delta\ell$ in the same basis. Using the definition of $\tilde x_i$ yields
\begin{equation}
  \tilde v_2 = \sum_i \tilde x_iv_2^{(i)} = \sum_i(1+\psi_{x_i}\delta\ell)x_iv_2^{(i)} = v_2 + \sum_i\psi_{x_i}x_i \sum_k a^{(i)}_k f_k\delta\ell \overset{!}{=} v_2 + I
\end{equation}
and thus by comparing the basis coefficients
\begin{equation}
  \sum_i a^{(i)}_k x_i\psi_{x_i} = b_k(\vec x)
\end{equation}
for all $k$. This is a linear system of equations for the graphical corrections $\psi_{x_i}$ determined by the $b_k$ and the vertex structure encoded in the coefficients $a^{(i)}_k$.


\end{document}